\documentclass{revtex4}
\usepackage{graphicx}% Include figure files
\usepackage{amsmath}
\usepackage{amssymb}
\usepackage{bbm}
\usepackage{}
\begin{document}

\title{Obtaining a scalar fifth force for matter via a conformal coupling to the scalar field}

\author{Hai-Chao Zhang}
\email{zhanghc@siom.ac.cn}

\affiliation{Key Laboratory for Quantum Optics, Shanghai Institute of Optics and Fine Mechanics, Chinese Academy of Sciences, Shanghai 201800, China }

\date{\today}

\begin{abstract}
In the framework of special relativity (SR), I propose that matter conformally couples to a scalar field through the Lagrangian density of matter, whether matter is characterized by classical or by quantum (statistical) mechanics. The largest interaction strength of the scalar-mediated force can achieve the order of $1/\Lambda_E^2$ with the cosmological constant $\Lambda_E\approx 2.4\,\rm{meV}$. This is about 60 orders of magnitude larger than Newtonian gravity, i.e., ${(M_{\mathrm{Pl}}/\Lambda_E)}^2\sim 10^{60}$, where $M_{\mathrm{Pl}}\approx2.4\times 10^{18}\, \rm{GeV} $ is the reduced Planck energy. However, the discrete $\mathbb{Z}_2$ symmetry of the conformal coupling does not allow a massless scalar field, satisfying the constraint that no evidence of the long range force is observed. The interaction range of the force is inversely proportional to the square root of the ambient density. Therefore, in the general situation in which the ambient density is larger than the current cosmic density, the interaction range is always smaller than $1/\Lambda_E \approx 80\, \mu\rm{m}$. The form of general relativity (GR) still holds even when the energy-momentum tensors of the quintessence and the conformal interaction are included. The main purpose here is to overcome the puzzle in \cite{hcz2}, i.e., which frame is physical.
\end{abstract}

\maketitle

\section{Introduction}\label{Intro}

Because the cosmological parameters are constrained at a sub-percent level \cite{planck,z1i,z27,z28}, it is firmly believed that dark energy and dark matter fill of the Universe, at least at the late-time. The existence of dark matter inferred from the cosmology challenges the standard model of fundamental particles (SM). The dark matter fields filling all of space, however, can naturally constitute a system of reference (SoR) in general relativity (GR). A dark matter field can be regarded as an infinite number of classical point bodies in space which can be used to mark the spacial coordinates, while the Compton frequencies of the dark matter field can be regarded as clocks to record time of the spacial positions. This meets the concept of the SoR \cite{landau} in GR as well as the original ideas of GR that there is no empty space having inertial property \cite{BD1,BD2}.

Dark energy, however, seems to be difficult to harmonize with GR. The cosmological constant $\Lambda_E\approx 2.4\,\rm{meV}$ \cite{planck}, neither its `natural' value $2.4\times 10^{18}\, \rm{GeV} $ of the reduced Planck energy \cite{z30,z31,z32,z29} nor another `natural' value of zero \cite{z2,rrc1}. Thus, modifying GR now becomes a broad topic \cite{3PRD1,3PRD2,3PRD3}. In brief, there are at least three reasons for introducing scalar fields into GR: to drive the cosmic acceleration at the late-time \cite{rrc1,rrc2}, to drive inflation of the early Universe \cite{KAO}, and to improve GR's compatibility with Mach's principle (MP) \cite{BD1,BD2}. Inflationary scalar field with a self-interaction potential density (SPD) which defines the scalar as quintessence, has been used to solve the flatness and horizon problems in the standard big-bang cosmology. However, questions are raised: after inflationary epoch, does the quintessence field really run out and disappear? Whether the inflationary quintessence field can also play the role of dark energy at the late-time. Many efforts have been made to use the matter-coupled quintessence to obtain the cosmological constant \cite{smc1,smc2,smc3,smc4,smc5} to drive the cosmic acceleration at the late-time.

To meet the requirement of Mp, Brans and Dicke in 1960s introduced a matter-coupled scalar field \cite{BD1,BD2}. Their scalar field does not possess SPD, namely, it lacks the basic characteristic of quintessence. The scalar field was introduced by the conformal transformation of metric tensors between the Einstein frame (EF) and the Jordan frame (JF). Although Brans-Dicke theory is still the most popular scheme in gravitational research, the way they introduced the scalar field has caused a lot of confusions. These confusions and controversies even led physicists to be roughly divided into six groups \cite{Bhadra,Faraoni}. The most fundamental confusion is that which frame is physical. This puzzle directly leads to the inability to use the scalar-tensor field theory to analyze experimental results. Physicists have to assume that a certain frame is physical to analyze a concrete experimental case. For example, in the laboratory, it is believed that the fifth force might be carried out in the EF \cite{z53}, while in astronomy, it is believed that the observation might be carried out in the JF \cite{NOGO}. In fact, I chose the EF to analyze our experiments \cite{hcz2}. Fortunately, the mathematical results of the experiments in that paper \cite{hcz2} are the same as that based on the technique in the current paper.

When quintessence field directly couples to ordinary matter (including dark matter), the scalar-mediated fifth force must appear. The direct coupling means that it is possible to analyze the scalar-mediated interaction without the GR framework, similar to the discussion of Coulombian force in electromagnetics. In order to introduce the new interaction in the SR framework, I assume that the quintessence is conformally coupled to matter through the Lagrangian density of matter. Obviously, one of the most important consequences of this procedure is that adding the matter-coupled quintessence would not change the form of GR. This is in line with Weinberg's theorem that a Lorentz invariant theory for spin-two-graviton must be GR \cite{wenb1,wenb2}. Therefore,the setup differs from \cite{Flanagan,hcz1,kaom,jdb,SC}. Considering that no evidence of the long range force is observed, I adopt a discrete symmetry to suppress long-range interactions. This is in sharp contrast with \cite{SC} in which the continuous symmetry is imposed.

In this paper, I show that the discovery of dark energy is supportive of the core idea of GR, i.e., general covariant and equivalence principle. In other words, the Einstein's field equations for the metric tensor together with the motion equation for the scalar quintenssence constitute GR. The introduction of quintessence strengthens rather than weakens GR. The reason is summarized as follows through mathematical formula.

In nonrelativistic mechanics, the Poisson equation
\begin{equation}
\nabla^2\varphi=4\pi G \rho \label{c1}
\end{equation}
for the gravitational field $\varphi$ satisfies the principle of equivalence. However, it does not possess Lorentz invariance. Namely, it does not satisfy the principle of relativity. This also means that the principle of equivalence does not necessarily lead to the mathematical form of the Einstein's field equations of GR. For example, to obtain the motion for the gravitational field under the constraints of both the principle of equivalence and the principle of relativity, one can use the d'Alembertian operator
\begin{equation}
\nabla^2-\frac{1}{c^2}\frac{\partial^2}{\partial t^2}\equiv\Box= \partial^2_\mu\label{s1}
\end{equation}
to replace the Laplacian operator $\nabla^2$ in Eq. (\ref{c1}) as long as the matter density $\rho$ is a scalar. The matter density $\rho$ is indeed a scalar in nonrelativistic mechanics. However, it is not a Lorentz invariant. In relativistic mechanics, $\rho$ corresponds to one of the components of the energy-momentum tensor $T^m_{\mu\nu}$ of matter, where index $m$ denotes matter. Therefore, using the trace $T^{m}$ of the energy-momentum tensor $T^m_{\mu\nu}$ instead of $\rho$, a Lorentz invariant equation of the gravitational field can be obtained as follows:
\begin{equation}
\Box \varphi= 4\pi G T^m. \label{s2}
\end{equation}
However, Eq. (\ref{s2}) cannot be used to explain the observable phenomena of the gravitational redshift and the deviation of light by larger stars since the property of $T^{\rm{photon}}=0$ indicates that the gravitational field does not couple to electromagnetic field.

Using Einstein's tensor $G_{\mu\nu}$ instead of $\varphi$ and $T^m_{\mu\nu}$ instead of $\rho$, Einstein obtained the Lorentz invariant equations to describe the usual gravitational interaction as follows:
\begin{equation}
G_{\mu\nu}={8\pi G}T^m_{\mu\nu}.\label{c2}
\end{equation}
The units $c=\hbar=1$ are adopted in this paper. The Einstein's field equations describe gravitational interaction very successfully. However, as pointed out by Dicke and Brans \cite{BD1,BD2}, a scalar field must be inserted into GR so as to meet the Mach's principle (MP).

Therefore, we need an improved motion equation for the scalar to replace Eq. (\ref{s2}) and a new Einstein's equations to replace Eq. (\ref{c2}) in the presence of the scalar. We also need the equations of motion for various kinds of matter in the presence of the scalar.

The article is organized as follows. In Sec. \ref{cc}, the conformal coupling is defined. In Sec. \ref{ifsr}, the equation of motion for the scalar field is obtained in the framework of special relativity. In Sec. \ref{eomforq}, the mathematical expressions of the sub-gravitational forces in the cases of the various forms of matter are described based on the SR framework. In Sec. \ref{Grresults}, the new Einstein's field equations in the presence of the quintessence are obtained based on the frame of GR. Then, the consequences of the quintessence field for the cosmology are explored. Finally, further discussion and the conclusion are shown in Sec. \ref{Concs}.

\section{Conformal coupling}\label{cc}

The equivalence of inertial and gravitational masses is the cornerstone of GR. In order to maintain the weak equivalence principle (WEP) when the new interaction is introduced, conformal coupling takes a simple form that various matter fields couple to the quintessence field $\phi$ via the same conformal fashion, i.e.,
\begin{equation}
{\mathcal{L}_I} = \mathcal{B}\left( \phi  \right){\mathcal{L}_m},\label{tq1}
\end{equation}
where $\mathcal{B}( \phi)$ is a dimensionless conformal coupling function of $\phi$, ${\mathcal{L}_m}$ is the Lorents-invariant Lagrangian density for any kind of matter fields (including the interaction terms between the various matter fields), respectively. Notice that we have now begun to work in the framework of SR.

Based on the Lagrangian density ${\mathcal{L}_m}$ in the absence of the quintessence field, the canonical energy-momentum tensor $T_{\mu \nu }^{(m)}$ of matter fields can be calculated \cite{Gupta,landau} [see also Eq. (\ref{3prd1018})]. From the relationship between the energy-momentum tensor and the Lagrangian density \cite{Gupta,landau}, the additional energy-momentum tensor $T_{\mu \nu }^{(I)}$ of matter and its trace $T_I$ in the presence of the quintessence field are related to the corresponding bare quantities $T_{\mu \nu }^{(m)}$ and $T_m$ via the same conformal fashion as follows:
\begin{subequations}\label{tq2}
\begin{eqnarray}
T_{\mu \nu }^{(I)} & = & \mathcal{B}\left( \phi  \right)T_{\mu \nu }^{(m)},\label{tq2a}\\
T_I & = & \mathcal{B}\left( \phi  \right)T_m.\label{tq2b}
\end{eqnarray}
\end{subequations}
Eq. (\ref{tq2b}) is an obvious corollary of Eq. (\ref{tq2a}). Eq. (\ref{tq2b}) is listed separately to highlight the coupling characteristics of electromagnetic field and quintessence field. Electromagnetic field has a property of the trace $T_m=0$ \cite{landau}. Thus, the corresponding $T_I$ also vanishes for the case of quintessence coupling to electromagnetic field, which does not mean that quintessence field has no effect on electromagnetic field. From Eq. (\ref{tq2a}), for example, the energy density of electromagnetic field can be changed by the quintessence field as long as $\mathcal{B}( \phi)\neq 0$.

\section{The equation of motion for the scalar field in the framework of special relativity }\label{ifsr}

For systems described by $n$ fields of $\varphi_r$ with $r=1, ..., n$, by varying the action $S=\int{\mathcal{L}}dVdt$ independently with respect to each field, the Euler-Lagrange equations for the $n$ fields can be obtained as follows:
\begin{equation}
\frac{\partial \mathcal{L}}{\partial\varphi_r}-\partial_\mu\frac{\partial\mathcal{L}}{\partial(\partial_\mu\varphi_r)}=0. \label{3prd10183}
\end{equation}
Consider the Lagrangian density for free quintessence field
\begin{equation}
{\mathcal{L}_\phi } =  - \frac{1}{2}({\partial _\mu }\phi )^2  - V(\phi ),\label{tq6}
\end{equation}
then the effective Lagrangian density for the quintessence field is the sum of ${\mathcal{L}_\phi } + {\mathcal{L}_I}$.  According to the Euler-Lagrange Eqs. (\ref{3prd10183}), the quintessence field equation of motion is given by:
\begin{equation}
\square \phi  = {V_{{,}\phi }}\left( \phi  \right) - {\mathcal{L}_m}{\mathcal{B}_{{,}\phi }}\left( \phi  \right),\label{tq7}
\end{equation}
where the subscript $``,\phi"$ denotes a partial derivative of $\partial/\partial \phi$. Eq. (\ref{tq7}) means that the scalar moves in an effective potential density (EPD) as follows:
\begin{equation}
{V_{{\text{eff}}}}\left( \phi  \right) = V\left( \phi  \right) - {\mathcal{L}_m}\mathcal{B}\left( \phi  \right).\label{tq8}
\end{equation}
One sees that the motion of the quintessence field is related to the Lagrangian density $ {\mathcal{L}_m}$ of matter rather than the trace $T_m$. Thus,  although $T_m=0$ for radiation, the motion of quintessence is directly influenced by radiation due to $ {\mathcal{L}_m}\neq 0$. This differs from \cite{hcz2}.

\subsection{A concrete form of conformal coupling}\label{acf}
Since there is no observed evidence that a long-range scalar field exists, we impose that the conformal coupling possesses a discrete $\mathbb{Z}_2$ spontaneously-broken symmetry \cite{LMK}and the SPD contains only $\phi^4$ term, i.e., \cite{hcz1}
\begin{subequations}\label{tq9}
\begin{eqnarray}
\mathcal{B}\left( \phi  \right) &=& \frac{1}{{4M_1^4}}{\left( {{\phi ^2} - M_2^2} \right)^2},\label{tq9a}\\
V(\phi ) &=& \frac{\lambda }{{\rm{4}}}{\phi ^4},\label{tq9b}
\end{eqnarray}
\end{subequations}
where ${M_1}$, ${M_2}$ and $\lambda $ are parameters of the model. In cosmology, the matter content (including radiation and dark matter) can be regarded as a set of perfect fluids. The Lagrangian density of a perfect fluid indexed by $i$ is the scalar of the proper density $\rho_i$, i.e., the energy in a local rest frame of the fluid. Thus the total Lagrangian density of all perfect fluids in the Universe is,
\begin{equation}
{\mathcal{L}_m} = -  \sum\limits_i \rho_i,\label{3prd0291}
\end{equation}
Based on the cosmological constraint, it has been demonstrated that in the most cosmic epochs (except for the inflationary era) \cite{hcz1}, at least after the era of the big bang nucleosynthesis (BBN) \cite{pbac}, the quintessence field must sit stably at the minimum $\phi_{\min}$ of the EPD. Substituting Eqs. (\ref{tq9}) and (\ref{3prd0291}) into Eq. (\ref{tq8}), the minimum of the EPD and the mass of the scalar field at the minimum are obtained as follows:
\begin{subequations}\label{tq11}
\begin{eqnarray}
\phi _{\min }^2 & = & \frac{{\rho M_2^2}}{{\lambda M_1^4 + \rho }} ,\label{tq11a} \\
m_{{\text{eff}}}^2 & \equiv & {V_{\rm{eff},\phi \phi }}({\phi _{\min }}) =  \frac{{2\rho M_2^2}}{{M_1^4}},\label{tq11b}
\end{eqnarray}
\end{subequations}
where $\rho  = \sum\nolimits_i {{\rho _i}} $ is the total energy density of the perfect fluids. The mass of the quintessence strongly depends on the matter density, leading to a short interaction range for large densities. However, the minimum of the EPD weakly depends on the matter density when $\rho  \gg \lambda M_1^4$, leading to the meaningful cosmological constant \cite{hcz1} in this setup. It should be emphasized that in the usual scalar-tensor theory \cite{hcz1} both the minimum and the mass of the scalar are also dependent of the pressure of the fluids.

The value of the SPD at the minimum can naturally act as the cosmological constant \cite{hcz1}, i.e., $\Lambda  = 8\pi G V\left( {{\phi_{\rm{min}}}} \right)$. However, in the framework of SR, it is inappropriate and indeed impossible to discuss the cosmic acceleration at late-time as well as the early cosmic inflation since they both involve the changes of the space-time itself. For example, the homogeneous cosmological constant generates a so-called ``repulsive force" on cosmic scale, which is very different from how the spatial gradient of the quintessence field causes the local fifth force shown as Eq. (\ref{tq4}). One has to work on the framework of GR to discuss the evolution of the Universe based on the sum of the energy-momentum tensors for all physical fields including the quintessence field.

\subsection{The quintessence field generated by a point mass}\label{qfg}

We consider a point mass embedded in the above perfect fluids, then the bare Lagrangian density for the system is
\begin{equation}
{\mathcal{L}_m} =  - \sum\limits_i {{\rho _i}}  - {M_C}\delta \left( {{\mathbf{r}} - {{\mathbf{r}}_C}} \right)\sqrt {1 - v_C^2} ,\label{tq12}
\end{equation}
where $M_C$, ${\mathbf{r}}_C$ and ${\mathbf{v}}_C$ are the mass, position vector and velocity of the particle, respectively. We can expand the field $\phi$ around the equilibrium value $\phi_{\min}$ as $\phi=\phi_{\min}+\delta\phi$. Notice that $\square \phi_{\min}=0$, one has
\begin{equation}
\left( \square   - {m_{{\text{eff}}}}^2 \right)\delta \phi  = {\mathcal{B}_{,\phi }}\left( {{\phi _{\min }}} \right){M_C}\delta \left( {{\mathbf{r}} - {{\mathbf{r}}_C}} \right)\sqrt {1 - v_C^2}.\label{tq14}
\end{equation}
One is always interested in a static, nonrelativistic case, i.e., the speed of light can be regarded as infinity, or $v_C=0$. Eq. (\ref{tq14}) is then rewritten as follows
\begin{equation}
\left( {{\nabla ^2} - m_b^2} \right)\delta \phi  = {\mathcal{B}_{,\phi }}\left( {{\phi _b}} \right){M_C}\delta \left( {{\mathbf{r}} - {{\mathbf{r}}_C}} \right),\label{tq15}
\end{equation}
in which we introduce symbols $\phi _b=\phi_{\min}$, $m_{\text{b}}=m_{{\text{eff}}}$ and $ \rho _b = \rho$ to highlight the background medium. The solution of Eq. (\ref{tq15}) is
\begin{equation}
\delta \phi  = -\frac{{{\mathcal{B}_{,\phi }} \left( {{\phi _b}} \right){M_C}}}{{4\pi r}}{e^{ - {m_{b}}r}}.\label{tq17}
\end{equation}
The acceleration of the test point mass $A$ influenced by the scalar $\delta \phi$ is obtained by Eq. (\ref{tq5plus}) as follows:
\begin{equation}
{\textit{\textbf{a}}_A} =  - \frac{{{\mathcal{B}_{,\phi }}\left( {{\phi _b}} \right)}}{{1 + \mathcal{B}\left( {{\phi _b}} \right)}}\nabla \delta \phi,\label{tq18}
\end{equation}
in which $ v_A\ll 1 $ has been used.
Substituting Eq. (\ref{tq17}) into Eq. (\ref{tq18}), one has
\begin{equation}
{\textit{\textbf{a}}_A} =  - {{\mathbf{e}}_r} {\frac{{\mathcal{B}^2_{,\phi }}\left( {{\phi _b}} \right)}{{{{1 + \mathcal{B}\left( {{\phi _b}} \right)}}}}}\left( {{m_b}r + 1} \right)\frac{{{M_C}}}{{4\pi r^2}} e^{  { - {m_b}r}}.\label{tq21}
\end{equation}
Therefore, the quintessence-mediated interaction range can be quantified by $m_b^{-1}$. The interaction strength can be approximately quantified by the coefficient ${{\mathcal{B}^2_{,\phi }}}({{{1 + \mathcal{B}}}})^{-1}\approx {\mathcal{B}^2_{,\phi }}\left( {{\phi _b}} \right)$, just like the role of the gravitational constant $G$ in Newtonian gravitational force. However, neither the interaction range nor the interaction strength is constant. They depend on the density of matter.

From Eq (\ref{tq11b}), considering a small background density of ${\rho _b} \ll \lambda M_1^4$, one has
\begin{subequations}\label{tq20}
\begin{eqnarray}
&& {\phi _b} \approx  \pm \frac{{{M_2}}}{{{M_1}^2}}{\left( {\frac{{{\rho _b}}}{\lambda }} \right)^{1/2}},\label{tq20a}  \\
&&  {\mathcal{B}_{,\phi }}\left( {{\phi _b}} \right) \approx  \mp \frac{{{M_2}^3}}{{{M_1}^6}}{\left( {\frac{{{\rho _b}}}{\lambda }} \right)^{1/2}} \ll \frac{{{M_2}^3}}{{{M_1}^4}}.\label{tq20b}
\end{eqnarray}
\end{subequations}
Thus, the interaction strength ${\mathcal{B}^2_{,\phi }}\left( {{\phi _b}} \right)$ is equal to zero when ${\rho _b} = 0$. Namely, no background medium, no sub-gravitational force. In fact, the quintessence field in general vanishes in the absence of matter, which is one of the key characteristics of the quintessence and is compatible with MP. For the large background density ${\rho _b} \gg \lambda M_1^4$, one has
\begin{subequations}\label{tq19}
\begin{eqnarray}
&&{\phi _b}\approx \pm {M_2}\left( {1 - \frac{{\lambda M_1^4}}{{2{\rho _b}}}} \right),\label{tq19a}\\
&& {\mathcal{B}_{,\phi }}\left( {{\phi _b}} \right) \approx \mp \frac{{\lambda M_2^3}}{{{\rho _b}}} \ll \frac{{{M_2}^3}}{{{M_1}^4}}.\label{tq19b}
\end{eqnarray}
\end{subequations}
Thus, the interaction strength ${\mathcal{B}^2_{,\phi }}\left( {{\phi _b}} \right)$ is also equal to zero when ${\rho _b}\to \infty$, although the quintessence ${\phi _b}=\pm M_2$ does not vanish. The largest value of ${\mathcal{B}^2_{,\phi }}=4{M_2}^6/(27{{M_1}^8})$ occurs at ${\rho _b} = \lambda M_1^4/2$. On both sides of the maximum of ${{\mathcal{B}^2_{,\phi }}\left( {{\phi _b}} \right)}$, no matter whether the ambient density becomes larger or smaller, the interaction strength will become weaker.

\section{Scalar-mediated force In the framework of special relativity}\label{eomforq}

\subsection{Scalar-mediated force for a point mass}\label{sgf}

For a single particle with rest mass $M_A$, the Lagrangian density is, \cite{landau}
\begin{equation}
{\mathcal{L}_m} = - M_A \delta \left( {{\mathbf{r}} - {{\mathbf{r}}_A}} \right)\sqrt {1 - v_A^2},\label{tq3}
\end{equation}
where $\textbf{v}_A $ is the velocity of the particle, $\delta \left( {{\mathbf{r}} - {{\mathbf{r}}_A}} \right)$ is Dirac $\delta$ function with position vector ${\mathbf{r}}_A$ of the particle. Then the Lagrangian function is, \cite{landau}
\begin{equation}
{L_m} \equiv \int {{\mathcal{L}_m}d^3 r }=- M_A \sqrt {1 - v_A^2},\label{tq3plus}
\end{equation}
where $d^3 r$ is the element of volume. Since the interaction Lagrangian function is ${L_I} = \int {{\mathcal{L}_I}d^3 r}$, noticing Eqs. (\ref{tq1}) and (\ref{tq3}), one has
\begin{equation}
{L_I}=-\mathcal{B}\left[ \phi({\mathbf{r}}_A) \right] M_A \sqrt {1 - v_A^2} =\mathcal{B}\left[ \phi({\mathbf{r}}_A) \right]{L_m}.\label{tq3plus2}
\end{equation}
The effective Lagrangian function $L_{\rm{eff}}$ for the particle is the sum of ${{L_m} + {L_I}} $. We assume that the scalar field itself does not depend on the velocity of the particle. According to the usual Euler-Lagrange equations
\begin{equation}
\frac{d}{dt}\frac{\partial L_{\rm{eff}}}{\partial \mathbf{v}_A}=\frac{\partial L_{\rm{eff}}}{\partial \mathbf{r}_A}, \label{3prd10182}
\end{equation}
we obtain the scalar-mediated fifth force as follows:
\begin{equation}
\frac{{d{\mathbf{p}_A}}}{{dt}} +\frac{d\ln \left[ {1 + \mathcal{B}\left( \phi({\mathbf{r}}_A)  \right)} \right]}{dt} {\mathbf{p}_A} =  -M_A \sqrt {1 - v_A^2}\nabla \ln \left[ {1 + \mathcal{B}\left( \phi({\mathbf{r}}_A)  \right)} \right],\label{tq4}
\end{equation}
where the momentum $\mathbf{p}_A$ is defined by
\begin{equation}
{{\mathbf{p}}_A} \equiv \frac{{\partial {L_m}}}{{\partial {{\mathbf{v}}_A}}} = \frac{{M_A{{\mathbf{v}}_A}}}{{\sqrt {1 - v_A^2} }},\label{tq5}
\end{equation}
and the spatial gradient refers to $\nabla\equiv \partial/\partial{{\mathbf{r}}_A}$ \cite{landau}. One is always interested in a static case that the temporal change of the scalar field can be neglected, i.e., $d\mathcal{B}/dt=0$. Substituting Eq. (\ref{tq5}) into Eq. (\ref{tq4}), the acceleration ${\textit{\textbf{a}}_A}={\dot{{\mathbf{v}}}_A}$ of the particle influenced by the scalar is obtained as follows:
\begin{subequations}\label{tq5plus}
\begin{eqnarray}
{\textit{\textbf{a}}_A} &= & - {\left( {1 - {v_A^2}} \right)^2}\nabla \ln \left[ {1 + \mathcal{B}\left( \phi  \right)} \right], \rm {for} \; {\mathbf{v}}_A \parallel {{\textit{\textbf{a}}}_A}, \label{tq5plusa}\\
{\textit{\textbf{a}}_A} &= & - {\left( {1 - {v_A^2}} \right)}\nabla \ln \left[ {1 + \mathcal{B}\left( \phi  \right)} \right], \; \rm {for} \; {\mathbf{v}}_A \perp {{\textit{\textbf{a}}}_A}. \label{tq5plusb}
\end{eqnarray}
\end{subequations}
The fifth force will vanish when $v_A$ approaches the speed of light. The expressions of acceleration in the two cases \cite{landau} are different. In the case of low velocity, Eqs. (\ref{tq5plus}) are almost the same as that of the usual scalar-tensor theory \cite{NOGO}. On the right-hand of Eq. (\ref{tq4}), the mass $M_A$ can be considered to be related to the so-called sub-gravitational property of the particle [see also $M_C$ in Eq. (\ref{tq21})]. In Eq. (\ref{tq5}), the same mass $M_A$ represents the inertia property. The expressions of acceleration shown by Eq. (\ref{tq5plus}) are independent of the rest mass of particles, which means that the conformal coupling does not change WEP. In other words, WEP is also valid for the case of particles moving in the sub-gravitational fields. However, the sub-gravitational acceleration depends on the velocity of the particle, which differs from that the gravitational acceleration in Newton's theory is independent of the velocity.

The quantum mechanical presentation for the point mass can be obtained by using the quantization method, such as the path integral quantization or the canonical quantization.

\subsection{Scalar-mediated interaction for matter fields}\label{sfmf}

Up to now, we have mainly focused on the point mass. However, matter usually appears in the form of various fields, such as, a spinor field, an electromagnetic field and a scalar field. 
For a spinor field, the Lagrangian density \cite{Gupta} is ${\mathcal{L}_m} = {\partial _\mu }\bar \psi {\gamma _\mu }\psi  - \kappa \bar \psi \psi $, where $\kappa$ is a positive constant or zero and denotes the Compton wave-number of the spinor field. The character $\kappa$ is introduced to distinguish it from the mass of the particles of the quantum field, because the Dirac field now has not been quantized by the quantum theory. It is not a quantum field here. It is also not a classical wave field since the spinor field has no classical wave counterpart. Noticing the interaction shown as Eq. (\ref{tq1}), the effective Lagrangian density ${\mathcal{L}_{\mathrm{eff}}}$ for the spinor field is the sum of ${\mathcal{L}_m } + {\mathcal{L}_I}$. The action for the spinor field has the form $S=\int{\mathcal{L}_{\mathrm{eff}}}dVdt$. According to the Euler-Lagrange Eqs. (\ref{3prd10183}), the equations of motion for the spinor fields are obtained as follows:
\begin{equation}
{\gamma _\mu }{\partial _\mu }\psi  + \kappa \psi  + {\gamma _\mu }{\partial _\mu }\ln \left[ {1 + \mathcal{B}\left( \phi  \right)} \right]\psi  = 0.\label{tq25}
\end{equation}
When the field is quantized by the quantum field theory, the rest masses of all particles of the quantum field are the same as $m\equiv\hbar\kappa/c=\kappa$ with the adaption of $\hbar=c=1$. Although the new summation terms ${\gamma _\mu }{\partial _\mu }\ln \left[ {1 + B\left( \phi  \right)} \right]\psi$ depend on the spinor field (operator), it is independent of the rest mass of the spinor field. This means that WEP is still valid in the presence of quintessence.

For a classical electromagnetic field, the Lagrangian density \cite{Gupta} is ${\mathcal{L}_m} =  - \frac{1}{4}{\left( {{F_{\mu \nu }}} \right)^2}$ with ${F_{\mu \nu }}=\partial_\mu A_\nu -\partial_\nu A_\mu$. Therefore, in the presence of the quintessence, the equations of motion for the electromagnetic fields are
\begin{equation}
{\partial _\mu }{F_{\mu \nu }} + {\partial _\mu }\ln \left[ {1 + B\left( \phi  \right)} \right]{F_{\mu \nu }} = 0.\label{tq23}
\end{equation}
Obviously, the motion of electromagnetic field is influenced by the quintessence. In stark contrast, electromagnetic field does not directly influence the motion of quintessence due to the trace $T_m=0$ for an electromagnetic field, which has been shown by Eq. (\ref{tq7}).

For a scalar field of matter, the Lagrangian density ${\mathcal{L}_m} =  - \frac{1}{2}{\left( {{\partial _\mu }\psi } \right)^2} - \frac{1}{2}{\kappa ^2}{\psi ^2}$, where $\kappa$ denotes the Compton wave-number of the scalar field \cite{Gupta}. Thus, in the presence of the quintessence, the equation of motion for the scalar is
\begin{equation}
\square \psi  - {\kappa ^2}\psi  + {\partial _\mu }\ln \left[ {1 + B\left( \phi  \right)} \right]{\partial _\mu }\psi  = 0.\label{tq24}
\end{equation}
Similar to Eq. (\ref{tq25}), when the scalar is quantized by the quantum field theory, $\kappa$ plays the role of the rest mass of the particles of the quantum scalar, i.e., $m=\kappa$. The rest-mass independence of the new summation terms ${\partial _\mu }\ln \left[ {1 + B\left( \phi  \right)} \right]{\partial _\mu }\psi $ means that the quintessence field does not influence WEP, although it indeed influences the motion of matter.

All in all, according to the Euler-Lagrange equations, the new summation terms represent the effect of the so-called sub-gravitational ``generalized force" on a given matter field. Each term in the summation is always proportional to the component of a four-dimensional gradient $ {\partial _\mu }\mathcal{B}\left( \phi  \right)$. Therefore, one can regard $ {\partial _\mu }\mathcal{B}\left( \phi  \right)$ as a certain kind of four-dimensional field intensity of the sub-gravity. The conformal coupling function $\mathcal{B}\left( \phi  \right)$ itself can then be regarded as the sub-gravitational potential generated by the quintessence field. However, since the summation terms also strongly involve the motion state of matter field, the sub-gravitational ``force" is fundamentally different from the gravitational force of Newton. Only in the case of point mass in classical physics can the sub-gravitational force be intuitively described. In any case, the existence of the so-called sub-gravitational potential indicates that the new sub-gravitational interaction does not influence WEP.

\section{the Einstein's field equations}\label{Grresults}

We now generalize our discussion from the cartesian coordinates in Sec. \ref{ifsr} to curvilinear coordinates and the the metric is ($- + +  + $). The quintessence field, matter, and the conformal interaction between the quintessence and matter are minimally coupled to gravitational field, i.e.,
\begin{equation}
S = \int {{d^4}x\sqrt { - g} } \left[ {\frac{1}{{16\pi G}}R+ {\mathcal{L}_m} +  \mathcal{B}\left( \phi  \right){\mathcal{L}_m}} + {\mathcal{L}_\phi } \right],\label{3prdgr1}
\end{equation}
where $R$ is the Ricci curvature scalar and $g$ is the determinant of the metric tensor $g_{\mu\nu}$. From the principle of least action, in the case of the presence of the quintessence field, the Einstein's equations can be obtained by varying the action (\ref{3prdgr1}) with respect to the metric as follows:
\begin{equation}
G_{\mu\nu}={8\pi G}\left( T^{(m)}_{\mu\nu} +  {\mathcal{B}\left( \phi  \right)}\sum\limits_m T^{(m)}_{\mu\nu} +T^{(\phi)}_{\mu\nu}\right),\label{c4}
\end{equation}
where $G_{\mu\nu}$ is the Einstein tensor, and
\begin{equation}
{T_{\mu \nu }} =  - \frac{2}{{\sqrt { - g} }}\frac{{\partial \left( {\sqrt { - g} \mathcal{L}} \right)}}{{\partial {g^{\mu \nu }}}} \label{3prd1018}
\end{equation}
defines the energy-momentum tensor. It is assumed that the coupling function does not depend explicitly on the metric and the Lagrangian density does not contain any derivatives of the metric. The equation of motion for the quintessence field in the presence of gravitational field is obtained by varying the action (\ref{3prdgr1}) with respect to $\phi$ as follows:
\begin{equation}
D^\mu D_\mu \phi=V_{,\phi} (\phi) +  {\mathcal{B}_{,\phi}\left( \phi  \right) }{\mathcal{L}_m},\label{3prd1n1}
\end{equation}
where $D_\mu$ is the covariant derivative with respect to the metric $g_{\mu\nu}$, and $D^\mu=g^{\mu\nu}D_\nu$.

It is evident, in the presence of the quintessence field, that the total energy-momentum tensor of matter plus the quintessence field must have a vanishing covariant divergence, i.e.,
\begin{equation}
\left( T^{(m)\nu}_{\mu} + {\mathcal{B}\left( \phi  \right)}T^{(m)\nu}_{\mu} +T^{(\phi)\nu}_{\mu}\right)_{;\nu}  = 0, \\ \label{3prdgr092923}
\end{equation}
where the subscript $``;\nu"$ denotes the covariant derivative $ D_{\nu}$. Both the energy-momentum tensor $T^{(\phi)}_{\mu\nu}$ of the quintessence field and $T^{(m)}_{\mu\nu}$ of matter are no longer conserved respectively due to the conformal coupling between the quintessence and matter.

\subsection{Cosmological evolution in the conformal quintessence field}\label{grpart2}

We consider a homogenous and isotropic cosmology with a scale factor $a(t)$ described by the line elements
\begin{equation}
d{s^2} =  - d{t^2} + {a^2}\left( t \right)\left[ {\frac{{d{r^2}}}{{1 - K{r^2}}} + {r^2}\left( {d{\theta ^2} + {{\sin }^2}\theta d{\phi ^2}} \right)} \right].\label{3prdgr2}
\end{equation}
where $K=1$, $0$, or $-1$ correspond to closed, flat, or open spaces, respectively. We will write out the equations of motion for quintessence and electromagnetic field in this so-called Friedmann-Roberson-Walker (FRW) metric. The The spaces contain several species of noninteracting perfect fluids of matter source. For a perfect fluid indexed by $i$ with Lagrangian density of ${\mathcal{L}_m} = -  \rho_i$, one deduces the energy-momentum tensor as follows:
\begin{equation}
T^{(m)}_{\mu\nu}=(\rho_i+p_i)u_{\mu}u_{\nu}+p_ig_{\mu \nu},\label{3prd1001}
\end{equation}
where $p_i$ is the pressure of the fluid.

Using Eqs. (\ref{3prd1n1}), (\ref{3prdgr2}) and (\ref{3prd1001}), the equations of motion for the quintessence and the perfect fluids are
\begin{subequations}\label{3prde1003}
\begin{eqnarray}
\ddot \phi  + 3H\dot \phi  + \frac{1}{a^2(t)}{\nabla^2_{\rm{FRW}}}\phi  + {V_{,\phi }}\left( \phi  \right) + {\mathcal{B}_{,\phi }}\left( \phi  \right)\sum\limits_i {{\rho _i}}  = 0, \label{3prde10031}\\
{{\dot \rho }_i} + 3H\left( {{\rho _i} + {p_i}} \right) = 0,\label{3prde10032}
\end{eqnarray}
\end{subequations}
respectively, where the overdots denote derivatives with respective to time, $H = \dot a/a$ is the Hubble parameter and ${\nabla^2_{\rm{FRW}}}$ is the FRW Laplacian operator defined by
\begin{equation}
{\nabla^2_{\rm{FRW}}} = \frac{{\sqrt {1 - K{r^2}} }}{{{r^2}}}\frac{\partial }{{\partial r}}\left( {{r^2}\sqrt {1 - K{r^2}} \frac{\partial }{{\partial r}}} \right) + \frac{1}{{{r^2}\sin \theta }}\frac{\partial }{{\partial \theta }}\left( {\sin \theta \frac{\partial }{{\partial \theta }}} \right) + \frac{1}{{{r^2}{{\sin }^2}\theta }}\frac{{{\partial ^2}}}{{\partial {\varphi ^2}}}.\label{3prde10131}
\end{equation}
Unlike the usual scalar-tensor theory \cite{hcz1,z5} in which the conservation law of the matter density in the EF is introduced by a definition, the conservation law of Eq. (\ref{3prde10032}) appears naturally here in the presence of the quintessence. One sees that Eq. (\ref{3prde10031}) describes a damped (negative damped \cite{note1}) oscillation of the quintessence for an expanding (a contracting) universe corresponding to $H>0$ ($H<0$). In the case of $H>0$, for example, with the time growing the quintessence eventually approaches the the minimum $\phi_{\rm{min}}$ of the EPD of the quintessence, i.e., ${V_{,\phi }}\left( \phi_{\rm{min}}  \right)+ {\mathcal{B}_{,\phi}\left( \phi_{\rm{min}}  \right) }  \sum\limits_i {{\rho _i}}  = 0$. It then sits stably at the minimum. Thus, in the equilibrium state, Eq. (\ref{3prde10031}) can be rewritten as follows:
\begin{subequations}\label{3prde1004}
\begin{eqnarray}
{V_{,\phi }}\left( {{\phi _b}} \right) + {\mathcal{B}_{,\phi }}\left( {{\phi _b}} \right){\rho _b} = 0, \label{3prde10041}\\
{\nabla^2_{\rm{FRW}}}{\phi _b} = 0,\label{3prde10042}\\
{{\ddot \phi }_b} + 3H{{\dot \phi }_b} = 0,\label{3prde10043}
\end{eqnarray}
\end{subequations}
in which we introduce symbols $ \rho _b = \sum\limits_i {{\rho _i}}$ and $\phi _b=\phi_{\min}$ to highlight the homogeneous background medium of matter and the equilibrium state of the quintessence.

In the equilibrium state of the quintessence, using Eqs. (\ref{c4}) and (\ref{3prdgr2}) we obtain the equations of the cosmic evolution
\begin{subequations}\label{3prde09301}
\begin{eqnarray}
{H^2} &\equiv & {\left( {\frac{{\dot a}}{a}} \right)^2}  = \frac{{8\pi G}}{3}\left( {V\left( {{\phi _b}} \right) + \frac{1}{2}{{\dot \phi }_b}^2 + \left( {\mathcal{B}\left( {{\phi _b}} \right) + 1} \right){\rho _b}} \right) - \frac{K}{{{a^2}}},\label{3prde093011}\\
\frac{{\ddot a}}{a} &=& \frac{{4\pi G}}{3}\left( {2V\left( {{\phi _b}} \right) - 2{{\dot \phi }_b}^2 - \left( {\mathcal{B}\left( \phi_b  \right) + 1} \right)\left( {{\rho _b} + 3{p_b}} \right)} \right),\label{3prde093012}
\end{eqnarray}
\end{subequations}
where ${p_b} = \sum\nolimits_i {{p_i}} $ denotes the total pressure of the fluids in the absence of the quintessence field. We compare Eqs. (\ref{3prde09301}) with  the $\Lambda$CDM model \cite{z26,z100} of
\begin{subequations}\label{3prde1011}
\begin{eqnarray}
{H^2} &=& \frac{\Lambda }{3} + \frac{{8\pi G}}{3}{\rho _{\text{m}}} - \frac{K}{{{a^2}}}, \label{equ10112}\\
\frac{{\ddot a}}{a}& =& \frac{{\Lambda }}{3} - \frac{{4\pi G}}{3}(\rho_{\rm{m}}+3p_{\rm{m}}),\label{equ13}
\end{eqnarray}
\end{subequations}
where $\Lambda$, $\rho_{\rm{m}}$ and $p_{\rm{m}}$ are the cosmological constant, the matter density and the pressure in the $\Lambda$CDM model, respectively. One can see that the value of the self-interaction potential at the minimum acts as the cosmological constant, i.e.,
\begin{equation}
\Lambda  = 8\pi G V( {{\phi_{b}}} ).\label{3prde09303}
\end{equation}
Also, the mass density and the pressure become now quintessence-dependent as follows:
\begin{subequations}\label{3prde1007}
\begin{eqnarray}
\rho_{\rm{m}}& = &{\rho_b} \left( {\mathcal{B}\left( {{\phi _b}} \right) + 1} \right),\label{equ14rep1}\\
p_{\rm{m}}& =& {p_b} \left( {\mathcal{B}\left( {{\phi _b}} \right) + 1} \right).\label{equ14rep2}
\end{eqnarray}
\end{subequations}
However, the additional terms of ${{\dot \phi }_b}^2 /2$ in Eq. (\ref{3prde093011}) and $2{{\dot \phi }_b}^2$ in Eq. (\ref{3prde093012}) have no corresponding physical quantities in the $\Lambda$CDM model. Roughly speaking, they correspond to the kinetic energy ${{\dot \phi }_b}^2 /2$ of the quintessence, and like ordinary matter, they possess the positive pressure property.
Therefore, the setup can repeat the results of the standard $\Lambda CDM$ model without the EF- and the JF-frame. By choosing the parameters of the self-interaction potential and the conformal coupling function shown as Eq. (\ref{tq9}), the cosmological constant can be obtained from Eq. (\ref{tq11a}) as follows:
\begin{equation}
\Lambda  = 2\pi G\lambda M_2^4{\left( {\frac{{{\rho _b}}}{{\lambda M_1^4 + {\rho _b}}}} \right)^2}.\label{3prde09304}
\end{equation}
When the parameters' production $\lambda M_1^4 $ is smaller than the current cosmic density, the parameters' production $\lambda M_2^4 $ mimic the cosmological constant. Obviously, one can only determine the two parameters' productions rather than the three separately parameters by the astronomical observations due to the homogeneous assumption of the Universe.

\subsection{Solving the Hubble tension based on the conformal quintessence field}\label{grpart3}

One of the most important techniques to obtain the information of the cosmic evolution is by measuring the electromagnetic radiations from distant galaxies or the cosmic microwave background (CMB). We will prove that the electromagnetic waves are damped by not only the Hubble fiction but also the temporal variation of the quintessence field. Thus, the amplitude of the received waves decays larger than one has ever expected and the frequencies decrease greater than one has expected. The loss of the energy of the electromagnetic field converts into gravitation by Hubble fiction and converts into the quintessence by the conformal coupling.

Consider a test signal which is emitted from atoms in a distant comoving galaxy. The signal then passes through the perfect fluids for a long distance and time. Finally, the signal is measured by a comoving local observer on our Earth. Since the source of the test signal is in the distant galaxy, one suppose the frequency of the signal is equal to that emitted from the same kind of atoms in the laboratory. Then the variations of the quantities of the signal during the long distance travel can be deduced by comparison methods. For example, by comparing the received frequency of the test signal with the measured frequency of the reference signal of the same kind atoms in the laboratory, one can determine the frequency shift of the test signal.

We assume that the Lagrangian densities of the signal and the homogeneous background are $\mathcal{L}_{\rm{test}}$ and $ -{\rho _b}$, respectively. The resulting Lagrangian of the measuring system is then given by the sum as follows:
\begin{equation}
{\mathcal{L}_m} = \mathcal{L}_{\rm{test}} - {\rho _b}. \label{grpart1004}
\end{equation}

In the presence of both the gravitational field and the quintessence field, the equations of motion of electromagnetic field can be obtained from that in the SR frame, shown as Eq. (\ref{tq23}), by using the covariant derivative instead of ordinary derivative. The corresponding Lagrangian density of the electromagnetic field in GR frame is
\begin{equation}
\mathcal{L}_{\rm{test}} =  - \frac{1}{4}{F_{\mu \nu }}{F^{\mu \nu }}. \label{3prd10052}
\end{equation}
Thus, the equations of the test signal under the metric shown in Eq. (\ref{3prdgr2}) are calculated as follows:
\begin{equation}
{{\dot F}^{0i}} + 3H{F^{0i}} + \frac{1}{{1 + \mathcal{B}}}\frac{{\partial \mathcal{B}}}{{\partial t}}{F^{0i}} =  - {\partial _j}{F^{ji}} - {C^i} - \frac{1}{{1 + \mathcal{B}}}{\partial _j}\mathcal{B}{F^{ji}}, \label{grpart1005}
\end{equation}
where ${F^{0i}}$ with $i=1,2,3$ are the time-space components of the electromagnetic field tensor, i.e., the components of electric field intensity, and ${F^{ji}} $ with $i,j = 1,2,3$ are the space-space components, i.e., the components of magnetic field intensity, and $C^i$ with $i=1,2,3$ are
\begin{subequations}\label{3prde1012}
\begin{eqnarray}
{C^1}& = &\cot \theta {F^{21}},\label{3prde10121}\\
{C^2}& =& \left( {\frac{{Kr}}{{1 - K{r^2}}} + \frac{2}{r}} \right){F^{12}},\label{3prde10122}\\
{C^3}& = & \left( {\frac{{Kr}}{{1 - K{r^2}}} + \frac{2}{r}} \right){F^{13}} + \cot \theta {F^{23}},\label{3prde10123}
\end{eqnarray}
\end{subequations}
respectively.

From left-hand side of Eq. (\ref{grpart1005}), one sees that the attenuation of the received signal comes from two sources: the Hubble expansion and the variation of the conformal coupling. Therefore, in order to get the true expansion rate of the Universe, one must subtract the contribution of the quintessence field from the total attenuation. We introduce an apparent Hubble parameter $H_{{\text{app}}}$ to present the total attenuation, i.e.,
\begin{equation}
{H_{{\text{app}}}} = H + \frac{1}{3}\frac{1}{{1 + \mathcal{B}\left( \phi  \right)}}\frac{{\partial \mathcal{B}\left( \phi  \right)}}{{\partial t}}. \label{3prd100522}
\end{equation}

For the further comprehension, we also discuss the case of a matter field being a scalar, i.e.,
\begin{equation}
\mathcal{L}_{\rm{test}} = - \frac{1}{2}{g^{\mu \nu }}{\partial _\mu }\psi {\partial _\nu }\psi  - \frac{1}{2}{\kappa ^2}{\psi ^2}. \label{3prd10012}
\end{equation}
By generalizing Eq. (\ref{tq24}) from SR frame to GR frome, the equation of motion of the scalar matter field can be obtained as follows:
\begin{equation}
\ddot \psi  + 3H\dot \psi  + \frac{{\partial \ln \left( {1 + \mathcal{B}} \right)}}{{\partial t}}\dot \psi  + {\kappa ^2}\psi  = \frac{1}{{{a^2}(t)}}\nabla _{{\text{FRW}}}^2\psi  - \frac{1}{{{a^2}(t)}}{\nabla _{{\text{FRW}}}}\psi  \cdot {\nabla _{{\text{FRW}}}}\ln \left( {1 + \mathcal{B}} \right), \label{3prd1006}
\end{equation}
where the FRW gradient operator is
\begin{equation}
{\nabla _{{\text{FRW}}}} = {{\mathbf{e}}_r}\sqrt {1 - K{r^2}} \frac{\partial }{{\partial r}} + {{\mathbf{e}}_\theta }\frac{1}{r}\frac{\partial }{{\partial \theta }} + {{\mathbf{e}}_\varphi }\frac{1}{{r\sin \theta }}\frac{\partial }{{\partial \varphi }}.\label{3prd10132}
\end{equation}
One can easily see that the effective Hubble parameter is the same as shown in Eq. (\ref{3prd100522}).
Using Eq. (\ref{3prde10032}), we can rewrite Eq. (\ref{3prd100522}) as follows:
\begin{equation}
{H_{{\text{app}}}} = H\left( {1 - \frac{{\rho_b  + p_b}}{{1 + \mathcal{B}\left( {{\phi _{{b}}}} \right)}}\frac{{\partial \mathcal{B}\left( {{\phi _{{b}}}} \right)}}{{\partial \rho_b }}} \right). \label{3prd10062}
\end{equation}
Inserting Eq. (\ref{tq11a}) into Eq. (\ref{tq9a}), we obtain the conformal coupling as follows:
\begin{equation}
\mathcal{B}\left( {{\phi _{b}}} \right) = \frac{1}{4}{\left( {\frac{{\lambda {M_1}^2{M_2}^2}}{{\lambda {M_1}^4 + \rho_b }}} \right)^2}. \label{3prd10061}
\end{equation}

\begin{figure}
\centering
\includegraphics[width=250pt]{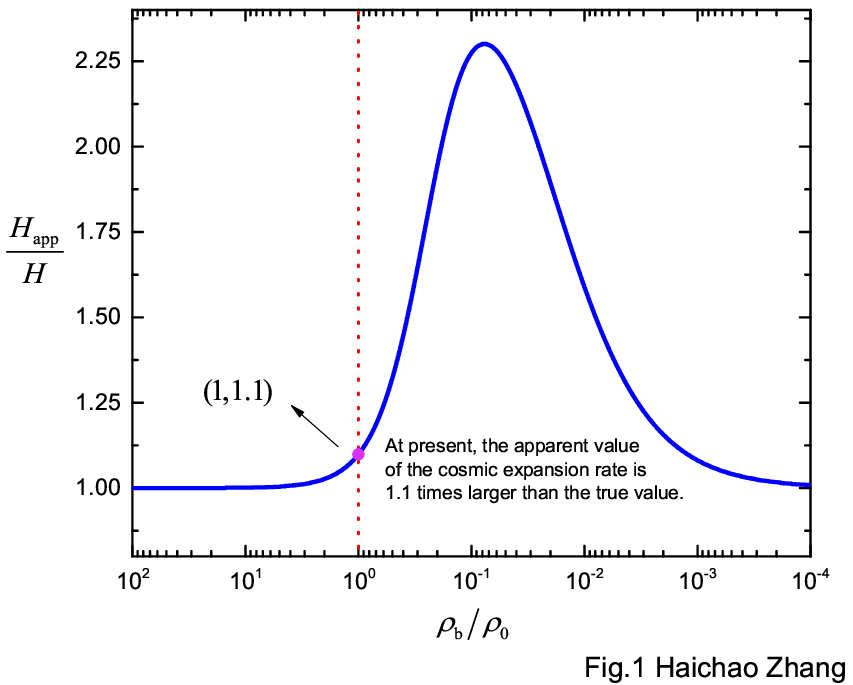}
\caption{The ratio of the apparent expansion rate of the Universe to the true expansion rate versus the cosmical density. The cosmical density is presented by the unit of the current matter density of the Universe.}\label{figure1}
\end{figure}

Figure \ref{figure1} shows the ratio of the apparent expansion rate of the Universe to the true expansion rate versus the cosmical density. In the past and the far future, the false increase of the cosmic expansion rate generated by the evolution of the scalar dark energy can be neglected. In the near future, the false increase of the cosmic expansion rate becomes bigger and bigger.

\section{Discussion and Conclusions}\label{Concs}

The orders of the parameters are constrained theoretically and experimentally as follows \cite{auw,cbj,hcz1,hcz2,planck}: $\lambda\sim\mathcal{O}(1) $, $M_2 \sim \mathcal{O}(\Lambda_E)$, $ M_1/M_2 \sim \mathcal{O}(1)$, and the current cosmic density $\sim \mathcal{O}(\Lambda^4_E)$ with $\Lambda_E\approx 2.4\,\rm{meV}$. Therefore, the maximum of ${{\mathcal{B}^2_{,\phi }}\left( {{\phi _b}} \right)}\sim \mathcal{O}(\Lambda_E^{-2})$ is greatly larger than $G=(8\pi M^2_{\mathrm{Pl}})^{-1}$, i.e., ${(M_{\mathrm{Pl}}/\Lambda_E)}^2\sim \mathcal{O}(10^{60})$, where $M_{\mathrm{Pl}}\approx2.4\times 10^{18}\, \rm{GeV} $ is the reduced Planck energy. At the maximum of ${{\mathcal{B}^2_{,\phi }}\left( {{\phi _b}} \right)}$, the interaction range is estimated from Eq. (\ref{tq11b}) to be $(\sqrt{\lambda} M_2)^{-1}\sim \mathcal{O}(\Lambda_E^{-1} )$ with $\Lambda_E^{-1}\approx 80\, \mu\rm{m}$. Due to $m_b^{-1} \propto \rho_b^{-1/2}$, one has the interaction range $m_b^{-1} \ll \Lambda_E^{-1}$ in the case of $\rho_b\gg \lambda M_1^4$. Thus, it is impossible to observe the sub-gravitational force in a very dense ambient density due to the very short interaction range.

When the Lagrangian density $\mathcal{L}_g$ for gravity \cite{landau} is included, the Einstein field equations and the equations of motion of matter and the quintessence field can be obtained by the usual variational principle. Since the quintessence is guessed to directly couple to matter through matter's Lagrangian density in a conformal way, our setup does not involve any statement of Jordan- and Einstein-frame that often appear in the usual scalar-tensor theory, such as Brans-Dicke's theory \cite{BD1,BD2}. The scalar field of Brans-Dicke's theory also does not possess the self-interaction potential density and then is not a quintessence field. However, Brans-Dicke's model in principle demonstrates that a gravitational theory compatible with MP should be characterized by a metric tensor together with a scalar field. We have shown above that the spatial gradient of the quintessence corresponds to a short-range sub-gravitational force based on the $\mathbb{Z}_2$ symmetry of the coupling function. Even if the spatial gradient of the quintessence field is zero in a matter-homogeneously-distributed universe, the self-interaction potential density of the quintessence would still generate a repulsive effect on the cosmic scale to drive the accelerated expansion of the Universe at late-time \cite{hcz1}. Since the evolution of the scalar dark energy can generate a false increase of the cosmic expansion rate, the setup provides a new method to solve the Hubble tension that the measured cosmic expansion rate is different between a direct and an indirect methodology.

\begin{acknowledgments}
I acknowledge discussions with Jie-Nian Zhang and Trent Claybaugh. This work was supported by the National Natural Science Foundation of China through Grant No. 12074396.
\end{acknowledgments}


\begin{thebibliography}{100}
\bibitem{note1} In a contracting universe of $H<0$, the quintessence eventually must depart away the equilibrium value $\phi_{\rm{min}}$ with the time growing. Thus, it no longer sits stably at the minimum but oscilates negatively-damped based on the positive feedback mechanism. However, in a relatively long time, the quintessence indeed had ever sat at the minimum. In this case, the value of the SPD at the minimum of can act as the cosmological constant.
\end{thebibliography}
\end{document}